\documentclass[iop]{emulateapj}

\def \mgii {Mg{\small~II}}

\def \civ {C{\small~IV}}

\def \feii {Fe{\small~II}}

\def \heii {He{\small~II}}

\def \hb {H~$\beta$}
\def \mbh {M_{\rm BH}}
\def \dks {D_{\rm KS}}
\def \ledd {L_{\rm Edd}}
\def \blr {_{\rm BLR}}
\def\kmsmpc {km$\,$s$^{-1}$Mpc$^{-1}$}
\def\om {\Omega_{\rm m}}
\def\ol {\Omega_{\Lambda}}
\def \ho {H_0}
\slugcomment{}

\shorttitle{Information from quasar broad emission lines}
\shortauthors{Croom}

\begin{document}


\title{Do quasar broad--line velocity widths add any information to
  virial black hole mass estimates?}


\author{Scott M. Croom\altaffilmark{1}}
\affil{Sydney Institute for Astronomy, School of Physics, University
  of Sydney, NSW 2006, Australia}


\altaffiltext{1}{email: scroom@physics.usyd.edu.au}


\begin{abstract}
We examine how much information measured broad--line widths add to
virial BH mass estimates for flux limited samples of quasars.  We do this
by comparing the BH masses estimates to those derived by randomly
reassigning the quasar broad--line widths to different objects and
re-calculating the BH mass.  

For 9000 BH masses derived from the \hb\ line we find that the
distributions of original and randomized BH masses in the
$\mbh$--redshift plane and the $\mbh$--luminosity plane are formally 
identical.  A 2D KS test does not find a difference at
$>90$\% confidence.  For the \mgii\ line (32000 quasars) we
do find very significant differences between the randomized and
original BH masses, but the amplitude of the difference is still
small. The difference for the \civ\ line (14000 quasars) is
$2-3\sigma$ and again the amplitude of the difference is small.
Subdividing the data into redshift and 
luminosity bins we find that the median absolute difference in BH mass
between the original and randomized data is 0.025, 0.01 and 0.04 dex
for \hb, \mgii\ and \civ\ respectively.  The maximum absolute
difference is always $\leq0.1$ dex.

We investigate whether our results are sensitive to corrections to \mgii\
virial masses, such as those suggested by Onken \& Kollmeier (2008).  These
corrections do not influence our results, other than to reduce the
significance of the difference between original and randomized BH
masses to only $1-2\sigma$ for \mgii.  Moreover, we demonstrate
that the correlation between mass residuals and Eddington ratio
discussed by Onken \& Kollmeier are more directly attributable to the
slope of the relation between \hb\ and \mgii\ line width.

The implication is that the measured quasar broad--line velocity widths
provide little extra information, after allowing for the mean velocity
width.  In this case virial estimates are equivalent to $\mbh\propto
L^{\alpha}$, with $L/\ledd\propto L^{1-\alpha}$ (with $\alpha\simeq0.5$).  This leaves an
unanswered question of why the accretion efficiency changes with
luminosity in just the right way to keep the mean broad--line widths
fixed as a function of luminosity.

\end{abstract}





\section{Introduction}

It is now clear that accretion onto super-massive black holes (BHs)
plays a crucial role in the process of galaxy formation and
evolution.  These BHs lie at the center of all massive galaxies and
are thought to influence galaxy evolution via powerful radiative
(i.e. quasar mode) and mechanical (radio mode) feedback (e.g. Croton
et al.\ 2006).  The tight correlation between BH mass and the
properties of the spheroidal component of their host galaxies
(e.g. Tremaine et al.\ 2002) infers that the growth of both is closely
related.  

The fundamental observables of black holes are mass and spin.  While
spin is still elusive (only being inferred from very indirect means),
there are now various approaches to the measurement of BH masses.  The
best constraint come from the BH at the centre of the Milky Way via
stellar kinematics (e.g. Gillessen et al.\ 2009) and the BH in NGC
4258 via mega-maser orbits (Miyoshi et al.\ 1995).  The masses of BHs
in other local galaxies can be estimated by the impact of
the BH on the motion of stars close to the nucleus, but model
degeneracies remain (Gebhardt \& Thomas 2009).  

For type-1 AGN, where we see all the way to the active nucleus, the
high velocity gas surrounding the BH gives us another probe of the BH
potential.  This is particularly powerful because it does not rely on
spatially resolving the sphere of influence of the BH and because the
the light from a quasar often overwhelms the light emanating from its
host galaxy.  As a result BH masses can be estimated in high redshift
quasars.  The disadvantage to this approach is that gas kinematics can
be more complex than the effectively collisionless stars.  Physical
processes such as radiation pressure and outflows (e.g. Marconi et
al.\ 2008) may cause potential biases in BH mass determinations,
although recent work suggests that this is not serious
(Netzer \& Marziani 2010). 

The approach to estimating BH masses using quasar broad emission lines
relies on the assumption that the gas is largely in virial motion
about the BH.  If this is the case, we then require an estimate of the
radius and velocity of the broad line region to estimate mass such that
\begin{equation}
\mbh=\frac{f\sigma\blr^2 R\blr}{G},
\label{eq:virial}
\end{equation}
where $f$ is a geometric factor, $R\blr$ is the radius of the broad
line region, $\sigma\blr^2$ is the velocity width of the line and $G$
is the gravitational constant.  The only viable approach to measuring
$R\blr$ is reverberation mapping (Blandford \& McKee 1983).  Evidence
from multiple lines within the same objects is consistent with virial
motion of the broad line region (Peterson \& Wandel 1999).
Reverberation mapping is limited to low redshift and low luminosity
broad line objects (i.e. Seyfert 1s), largely because time-dilation
and the scaling of $R\blr$ with luminosity (typically
$R\blr\sim L^{0.5}$ if driven by photo-ionization) can increase by an order of
magnitude or more the time lag between continuum and line
variability.  In order to make estimates of BH mass in higher redshift
quasars the correlation between $R\blr$ and luminosity found via
reverberation mapping (e.g. Kaspi et al.\ 2000) has been applied.
Using the $R\blr-L$ relation and calibrating to the \hb\ emission line
visible at low redshift, both the \mgii\ (McLure \& Jarvis 2002) and
\civ\ (Vestergaard et al.\ 2006) emission lines have been used for BH
mass estimation.  This approach, using the width of a broad optical or
UV emission line and the  $R\blr-L$ relation is generally termed the
{\it virial method}.  The power of such techniques is that they can be
applied to large optical spectroscopic samples of
quasars such as the Sloan Digital Sky Survey (SDSS; Schneider et al.\
2007), 2dF QSO Redshift Survey (2QZ; Croom et al.\ 2004) and 2dF-SDSS
LRG and QSO Survey (2SLAQ; Croom et al.\ 2009).  It is then in
principle possible to study properties such as mass functions
(e.g. Vestergaard et al.\ 2008) or the clustering of quasars as a
function of BH mass (e.g. Shen et al.\ 2009).

The above approaches give us great hope of characterizing the
super-massive BH population over most of cosmic time.  However, a
number key issues have been raised.  The \civ\ line in particular has
been a subject of close scrutiny (e.g. Netzer et al.\ 2007; Baskin \&
Laor 2005) largely due to the strong asymmetries and absorption often
seen in this line.  However some authors argue that \civ\ can be used
to provide a viable mass estimate (e.g. Vestergaard \& Peterson
2006).  This is because the asymmetries in \civ\ cause mass differences which are smaller than the
relatively large scatter in the virial estimates.  Vestergaard \&
Peterson also argue that for the handful of objects which have
reverberation mapping of multiple lines (including \civ), these
provide consistent BH mass estimates.  Not withstanding these points,
the \civ\ line is thought to be located much closer to the 
nucleus than \hb\ or \mgii\ so it is quite possible that different
dynamics are in play in these two regimes.

More generally, one of the recent challenges to the virial method is
the study of the dispersion in quasar broad line widths as a function
of luminosity for \mgii\ and \civ\ by Fine et al.\ (2008) and Fine et
al.\ (2010) respectively.  The most striking result from this work is
the small scatter in line widths for the most luminous quasars, which
is less than 0.1 dex for both the \mgii\ and \civ\ lines.  This
implies a scatter of less than 0.2 dex in BH mass.  Given that this
scatter must include dispersion due to accretion rate variations,
orientation effects, scatter in the $R\blr-L$ relation and
observational uncertainty, it is hard to see how such a small scatter
can be produced.  This scatter is also smaller than the typical
quoted uncertainties on virial BH mass estimates.  

In this paper we consider the low scatter present in broad line
velocity widths and address the question of how much information they
actually provide concerning virial BH mass estimates.  Our approach is
to carry out a simple test of re-determining BH masses after randomly
reassigning the broad-line velocity widths to different quasars.
Throughout this paper we assume a flat cosmology with $\ho=70$\kmsmpc,
$\om=0.3$ and $\ol=0.7$.

\section{Black hole masses}\label{sec:bhmass}

\begin{figure}
\hspace{-0.5cm}\includegraphics[scale=0.45]{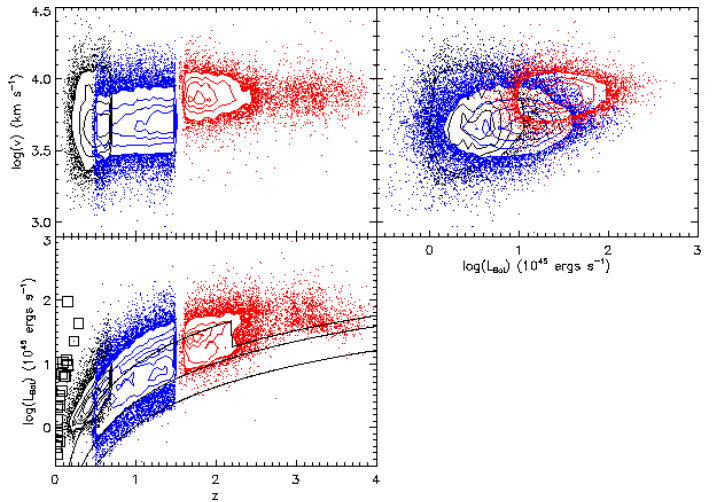}
\caption{The distribution of line width [$\log(v)$], bolometric
  luminosity [$\log(L_{\rm Bol})$] and redshift for the
  samples analysed in this paper.  The \hb\ (black), \mgii\ (blue) and
\civ\ (red) samples are shown separately.  Contours are linearly
spaced in 5 equal intervals.  The lower left plot also shows the AGN
with reverberation mapping used to calibrate virial relations [open
squares; taken from vestergaard \& Peterson (2006)] and the
approximate flux limits of the SDSS (top), 2QZ (middle) and 2SLAQ
(bottom) using the mean bolometric correction from the B-band given by McLure \&
Dunlop (2004).  For the 2QZ and 2SLAQ surveys the edge of the locus of
sources lies close to these limits, while for SDSS which is $i$-band
limited the sources have greater scatter across the nominal survey
limits.  \mgii\ and \civ\ line widths have been corrected to match the
FWHM measurements of the \hb\ lines assuming the lines are single
Gaussians.}
\label{fig:lvz}
\end{figure}

\begin{figure*}
\begin{centering}
\includegraphics[angle=270,scale=0.60]{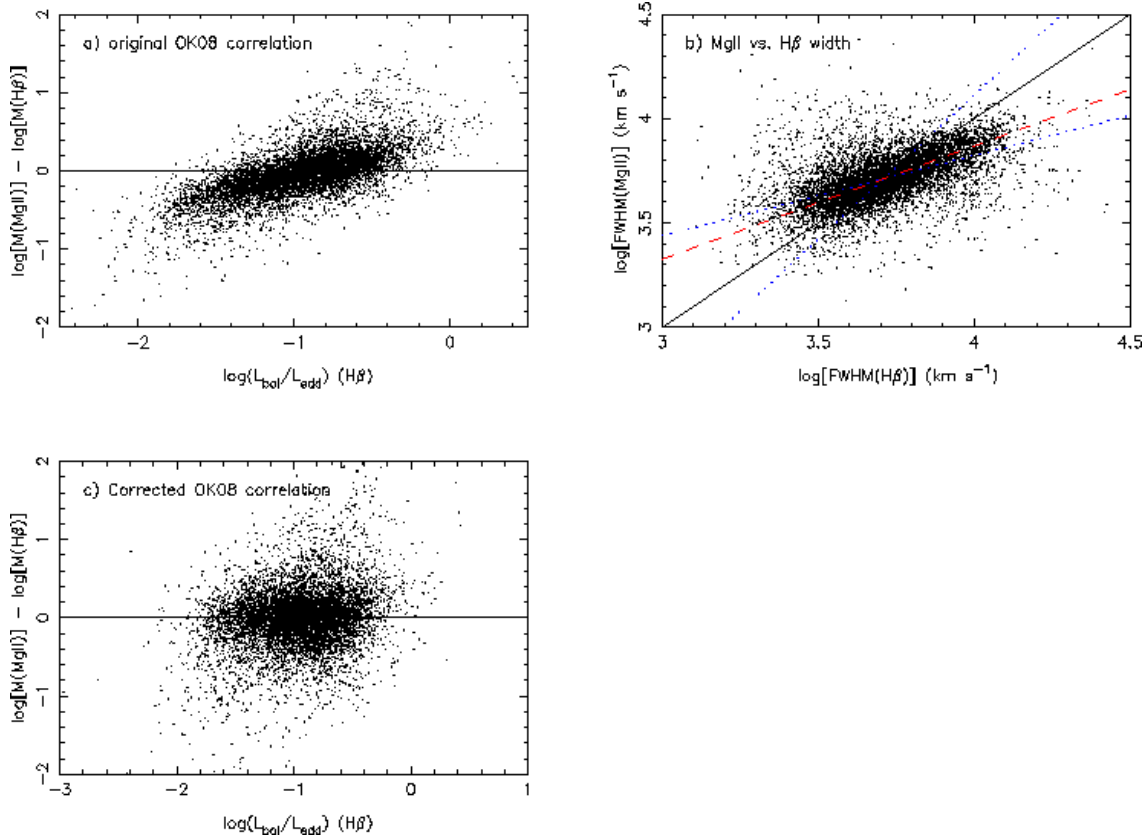}
\caption{a) The correlation between BH mass difference and Eddington
  ratio as given by OK08 using the data of Shen et al. (2008).  b) The
  correlation between \mgii\ and \hb\ velocity width from the same
  data, which shows a gradient which deviates from 1-to-1 (black solid
  line).  The OLS$(X|Y)$ and OLS$(Y|X)$ fits are shown by the blue
  dotted lines and the bi-sector of these is given by the red dashed
  line.  c) The relation between mass difference and Eddington ratio
  after correcting the \mgii\ velocity.}
\label{fig:ok}
\end{centering}
\end{figure*}


The measurements of \hb\ broad line widths is taken from Shen et
al.\ (2008), who use the fifth data release of the  SDSS quasar survey
(Schneider et al 2007).  Shen et al.\ largely follow the procedure of
McLure \& Dunlop (2004), including their calibration of the
virial BH mass estimator.  This involves first fitting the underlying
power--law continuum plus iron emission, which is then subtracted off.
They then fit a double Gaussian to the emission line, constraining one
component to be broad, and the other narrow.  The FWHM of the line is
then taken to be that of the broad Gaussian component.     

For the \mgii\ and \civ\ emission line fits
we use the results of Fine et al.\ (2008) and Fine et al.\ (2010)
respectively.  The main reason for using the Fine et al.\ data sets is
that as well as using quasars from the SDSS, they also use spectra
from the 2dF QSO Redshift Survey (2QZ; Croom et al. 2004) and the
2dF-SDSS LRG and QSO Survey (2SLAQ; Croom et al. 2009).  This provides
a larger dynamic range in luminosity at any given redshift.  The line
width measurements made by Fine et al.\ use the inter--percentile
value (IPV) width (e.g. Whittle 1985).  Continuum plus \feii\ emission
is first subtracted, and then a single Gaussian fitted.  The IPV width
is then measured over the range $\pm1.5\times$ the Gaussian FWHM.  The
IPV approach has the advantages that it is doesn't depend on a
particular line shape, and as it is derived from the cumulative sum
of flux in the line, easier quantification of the errors are
possible (particularly incorporating the covariance due to continuum
subtraction, which is generally not included in line fits).  In the case of \civ\ Fine et al.\ (2010) also fit and
subtract the contaminating \heii\ and {O{\small~III}]}\ at
  $\simeq1600$\AA.  The virial BH mass estimates use the McLure \&
  Dunlop (2004) and Vestergaard \& Peterson (2006) calibrations for \mgii\
  and \civ\ respectively.  The distribution of redshift, bolometric
  luminosity and linewidth is shown is Fig. \ref{fig:lvz}.  The
  fainter 2QZ and 2SLAQ samples allow us to probe a broader range of
  luminosities than just the SDSS alone, similar to the work of
  Kollmeier et al.\ (2006) who also target faint AGN (limited to
  $R<21.5$).

The exact form of the virial relation and its calibration is not
crucial in the analysis presented below.  Our aim is to determine how
much information the broad line velocity width contributes to
distribution of measured black hole masses.  To this end we perform
the simple test of randomizing the measured velocities.  That is,
assigning the velocity width of one quasar to another.  We do this
without reference to an object's redshift or luminosity (other than to
only randomize \hb\ widths with other \hb\ widths, and similarly for
\mgii\ and \civ), so that a quasar at the low luminosity and low
redshift end of the sample could be given a width from a high
redshift, high luminosity object.  If the
velocity widths are adding useful information to our black hole mass
estimates, then the distribution of {\it original} black hole masses
should be significantly different from the {\it randomized} black hole
masses.

\begin{figure*}
\includegraphics[scale=.45]{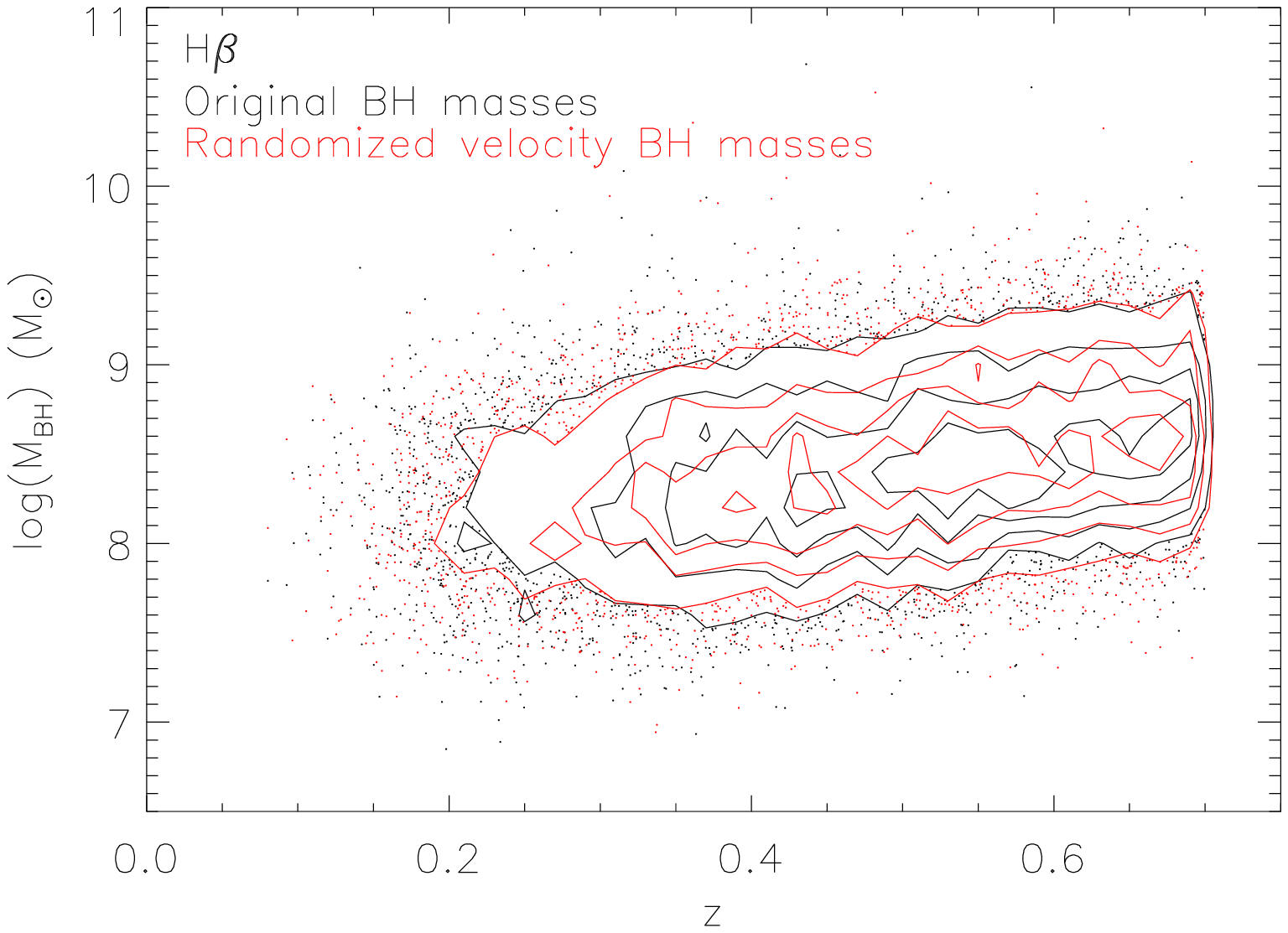}\includegraphics[scale=.45]{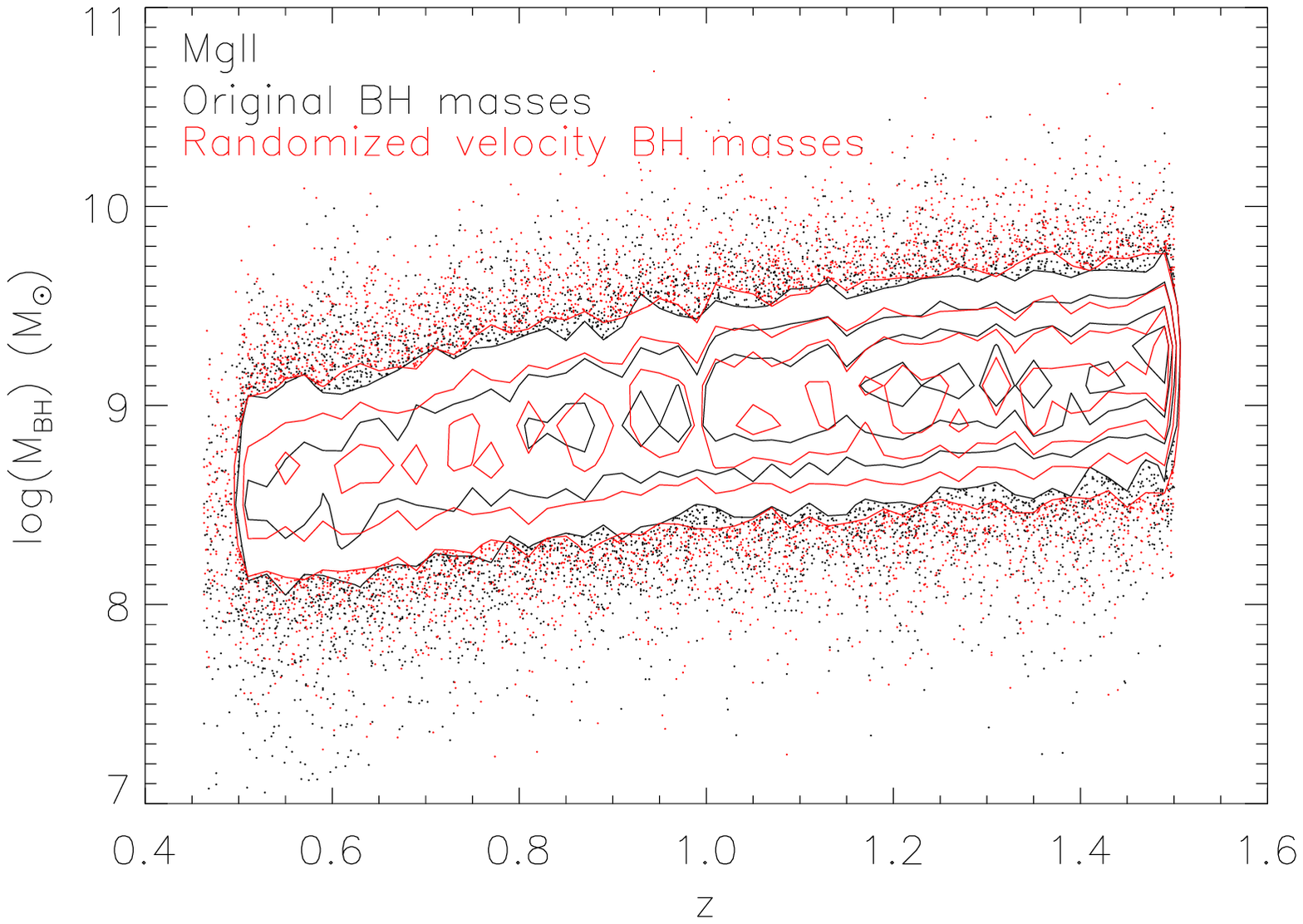}
\includegraphics[scale=.45]{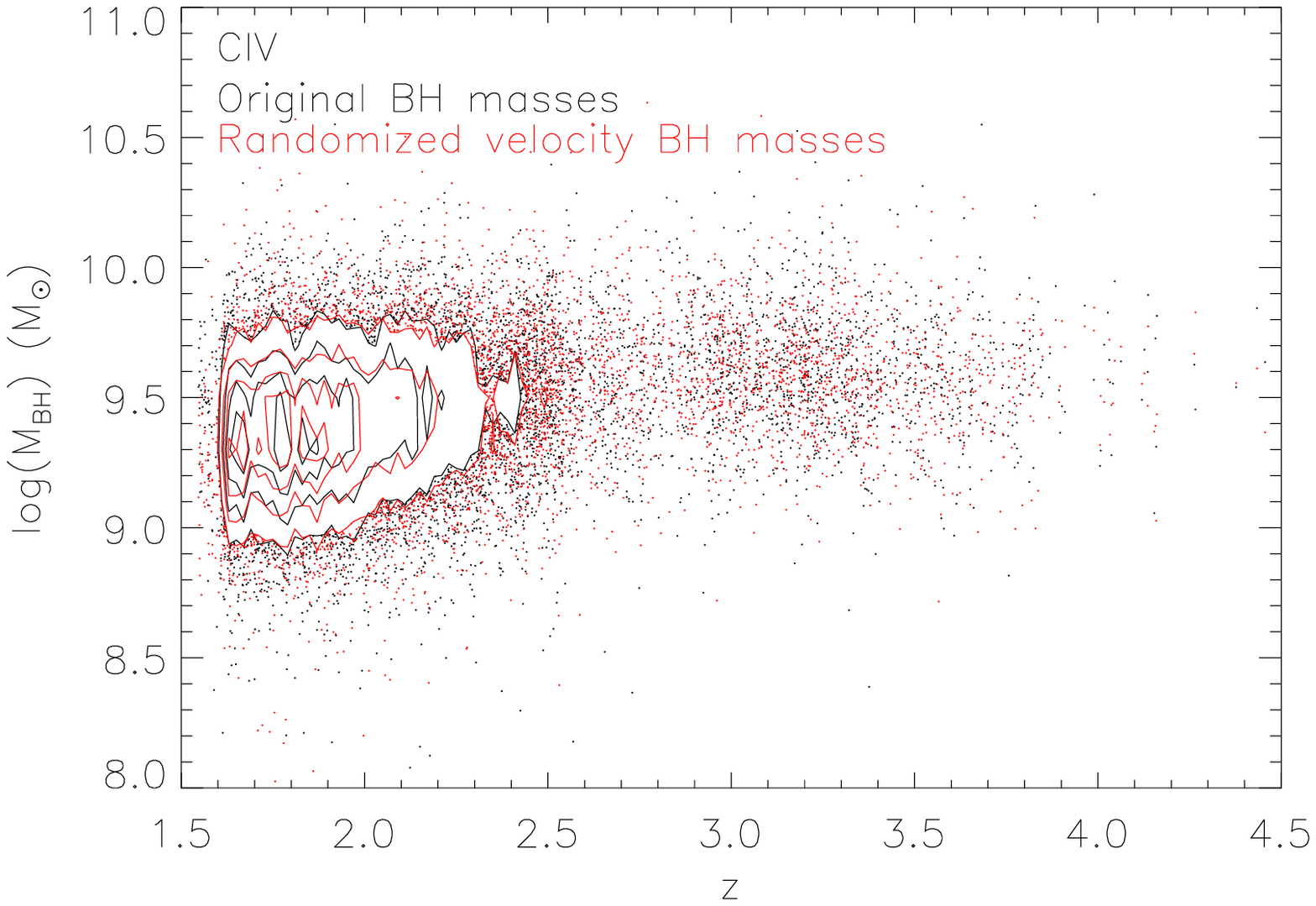}
\caption{The distribution of black hole mass estimates vs. redshift
  using the \hb\ (top left), \mgii\ (top right) and \civ\ (bottom left)
  lines.  The black contours and points are the original
  BH masses (Shen et al. 2008 for \hb\, Fine et al., 2008 for \mgii\
  and Fine et al. 2010 for \civ), while the red contours and points with
  the broad line velocities randomized.  The contours are used in regions
  of high density and are spaced at five equal linear intervals, using
  the same levels for both black and red contours.} 
\label{fig:hb}
\end{figure*}

The work of Onken \& Kollmeier (2008, OK08) suggested that \mgii\ based
virial methods contain a bias which is a function of Eddington ratio.
We also investigate whether such a bias influences our results
presented below.  OK08 present the bias as a mass
difference [$M$(\mgii)-$M$(\hb)] as a function of $L/\ledd$
(Fig. \ref{fig:ok}a).  However, these are both derived quantities
which are correlated in non-trivial ways.  In terms of measured (rather
than derived) quantities this bias is due to the relation between \hb\
and \mgii\ velocity width not having a gradient of 1 (see
fig. \ref{fig:ok}b).  OK08 define a matrix of corrections for
the \mgii\ mass estimates. However, we find that a much more direct
correction of the \mgii\ velocity widths also removes the bias.  We
find this correction by making an ordinary least squares (OLS) fit to
FWHM(\mgii) vs. FWHM (\hb) in both $(X|Y)$ and $(Y|X)$ and taking the
bisector (red dashed line in Fig. \ref{fig:ok}b; Isobe et al. 1990).
The resulting fit is
\begin{equation}
\log[{\rm FWHM(MgII)}]=1.70+0.54\log[{\rm FWHM(H\beta)}].
\label{eq:mgiicorr}
\end{equation}
Applying this relation to correct the \mgii\ velocity widths results
in an offset with respect to the original virial relation, which is
corrected by using
\begin{eqnarray}
\log\left(\frac{\mbh}{M_\odot}\right) & = & 0.505+0.62\log\left(\frac{\lambda L_\lambda}{10^{44}\rm{ergs~s^{-1}}}\right)\nonumber\\
                                     &   & +2\log\left(\frac{{\rm FWHM(MgII)_{\rm c}}}{{\rm km~s^{-1}}}\right)+1.278,
\label{eq:newmgiimass}
\end{eqnarray}

where the final term ($+1.278$) corrects the offset caused by the
\mgii\ velocity correction.  ${\rm FWHM(MgII)_{\rm c}}$ is the
\mgii\ width corrected to match \hb\ using Eq. \ref{eq:mgiicorr}.  The
result of applying the above BH mass
estimate, including the correction to \mgii\ velocities, is shown in
Fig. \ref{fig:ok}c.  It can be seen that this corrects the strong bias
originally presented by OK08.  Applying such a correction as part of
our analysis has no impact on the result presented in this paper.
Further details will be described below.  A gradient in FWHM(\mgii)
vs. FWHM (\hb) which deviates from one also has more broad
implications for the virial technique.  Assuming that the \hb\ line is
virialized and that the exponent for the velocity is 2 (as in Eq. \ref{eq:virial}), then the
implied exponent for the \mgii\ velocities is $2/0.54\simeq3.7$.  This
deviation from a naive virial relation is similar to the results of
Wang et al.\ (2009) who look at the calibration of virial relations
without the assumption that the exponent for the velocity is 2.  It
could be that the virial assumption is wrong for either \hb\ or \mgii\
or that the relative distribution of gas is markedly different.
Infact, \hb\ is expected to be due to multiple components
(e.g. Sulentic et al. 2006).  Our lack of understanding of the
differences between \hb\ and \mgii\ is a concern for the virial
approach. 

\begin{figure*}
\includegraphics[angle=270,scale=.65]{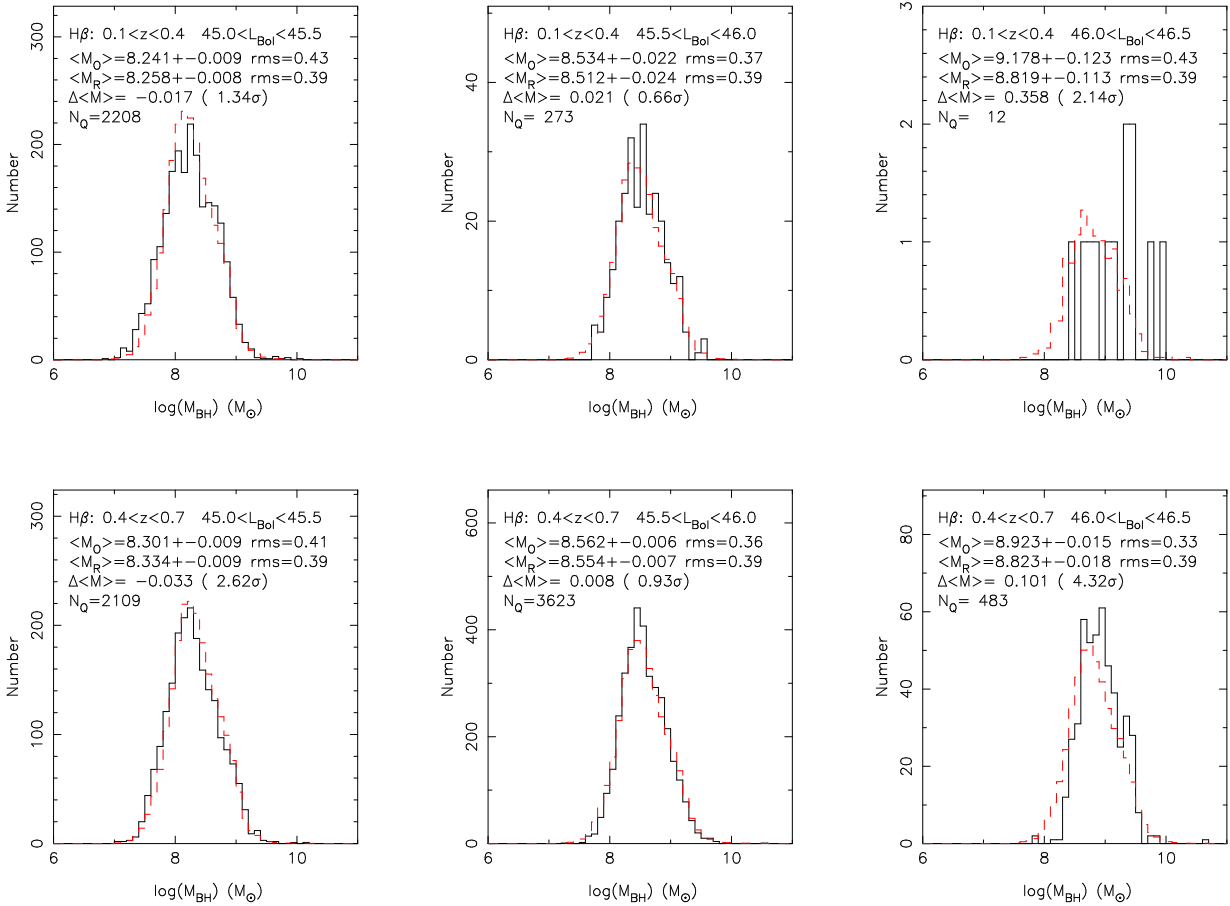}
\caption{The distribution of black hole mass estimates in different
  redshift and bolometric luminosity intervals for \hb.  We compare the
  original mass estimates ($M_{\rm O}$; black solid lines) to the
  randomized estimates ($M_{\rm R}$; red dashed lines).  Mean masses
  are given for both, along with the difference in the mass and its
  significance.}
\label{fig:hbhist}
\end{figure*}

\section{Results}

\begin{deluxetable}{llrrr}
\tablecaption{The results of a 2-D KS test comparing original and
  randomized BH masses in both the $\mbh-z$ and $\mbh-L$ planes.\label{tab:ks}}
\tablewidth{0pt}
\tablehead{
\colhead{Line} & \colhead{Test} & \colhead{N$_{\rm Q}$\tablenotemark{a}} & \colhead{$\dks$} & \colhead{$P(>D)$} 
}
\startdata
\hb\ (SDSS)  & $\mbh-z$ & 8979  & 0.0194 & 2.18E-01\\ 
\hb\ (SDSS)  & $\mbh-L$ & 8979  & 0.0223 & 9.38E-02\\ 
\mgii\ (all) & $\mbh-z$ & 32214 & 0.0290 & 2.94E-08\\ 
\mgii\ (all) & $\mbh-L$ & 32214 & 0.0230 & 8.27E-06\\ 
\mgii\ (all, corrected)\tablenotemark{b} & $\mbh-z$ & 32214 & 0.0280 & 7.32E-02\\ 
\mgii\ (all, corrected)\tablenotemark{b} & $\mbh-L$ & 32214 & 0.0236 & 1.79E-01\\ 
\mgii\ (SDSS) & $\mbh-z$ & 22910 & 0.0295 & 6.22E-06\\ 
\mgii\ (SDSS) & $\mbh-L$ & 22910 & 0.0259 & 8.10E-05\\ 
\mgii\ (2QZ) & $\mbh-z$ & 6784 & 0.0221 & 2.41E-01\\ 
\mgii\ (2QZ) & $\mbh-L$ & 6784 & 0.0212 & 2.54E-01\\ 
\mgii\ (2SLAQ) & $\mbh-z$ & 2491 & 0.0252 & 6.54E-01\\ 
\mgii\ (2SLAQ) & $\mbh-L$ & 2491 & 0.0278 & 5.21E-01\\ 
\civ\  (all) & $\mbh-z$ & 13795 & 0.0213 & 3.30E-02\\ 
\civ\  (all) & $\mbh-L$ & 13795 & 0.0257 & 2.33E-03\\ 
\civ\  (SDSS) & $\mbh-z$ & 11861 & 0.0215 & 5.97E-02\\ 
\civ\  (SDSS) & $\mbh-L$ & 11861 & 0.0255 & 9.94E-03\\ 
\civ\  (2QZ) & $\mbh-z$ & 1593 & 0.0341 & 5.71E-01\\ 
\civ\  (2QZ) & $\mbh-L$ & 1593 & 0.0303 & 6.78E-01\\ 
\civ\  (2SLAQ) & $\mbh-z$ & 307 & 0.0642 & 7.48E-01\\ 
\civ\  (2SLAQ) & $\mbh-L$ & 307 & 0.0572 & 8.24E-01\\ 
\enddata
\tablenotetext{a}{Number of quasars in each sample.}
\tablenotetext{b}{\mgii\ velocities and BH masses corrected as
  described in Eqs. \ref{eq:mgiicorr} and \ref{eq:newmgiimass}.}
\end{deluxetable}

\begin{figure*}
\includegraphics[angle=270,scale=.65]{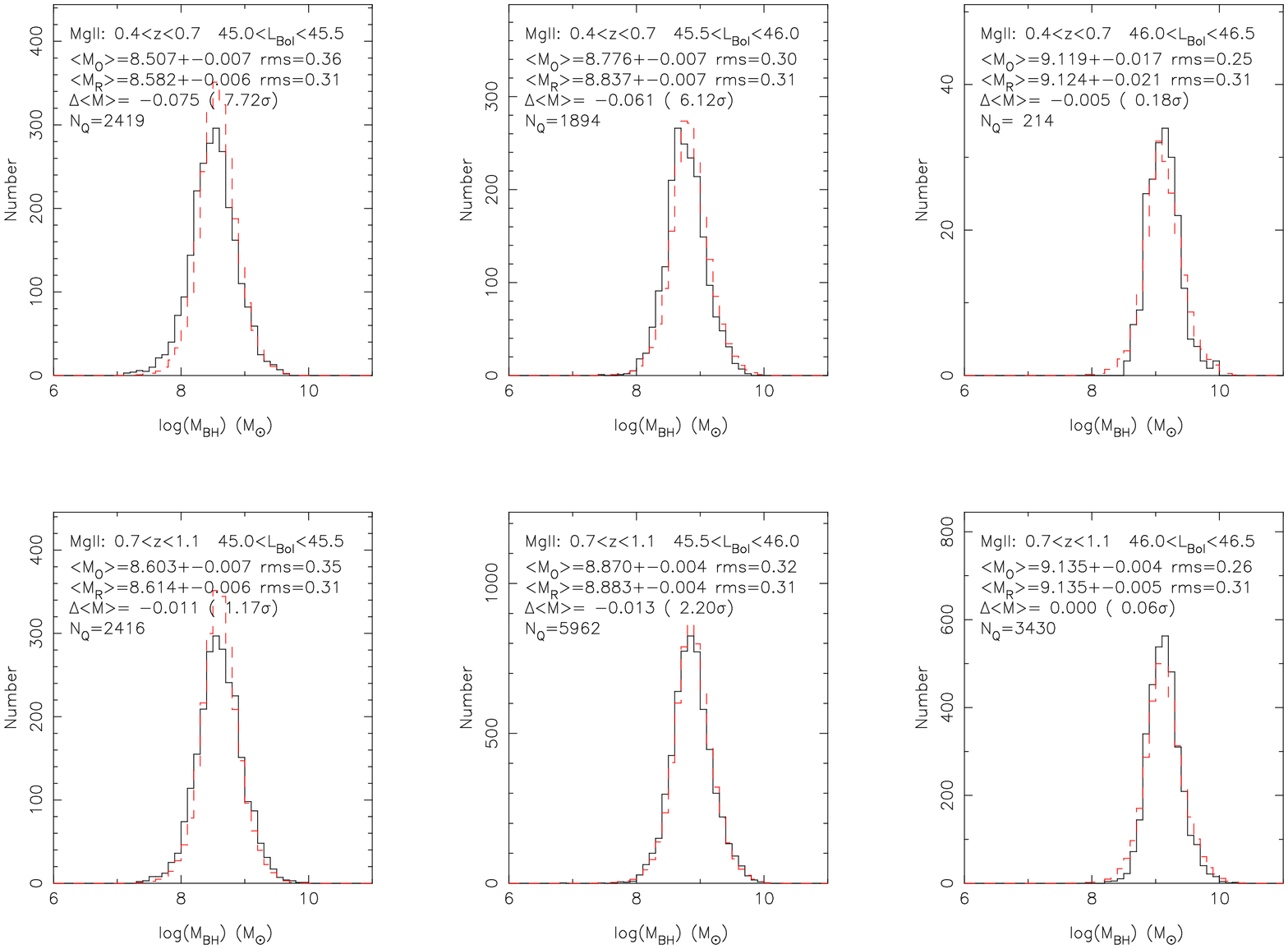}
\includegraphics[angle=270,scale=.65]{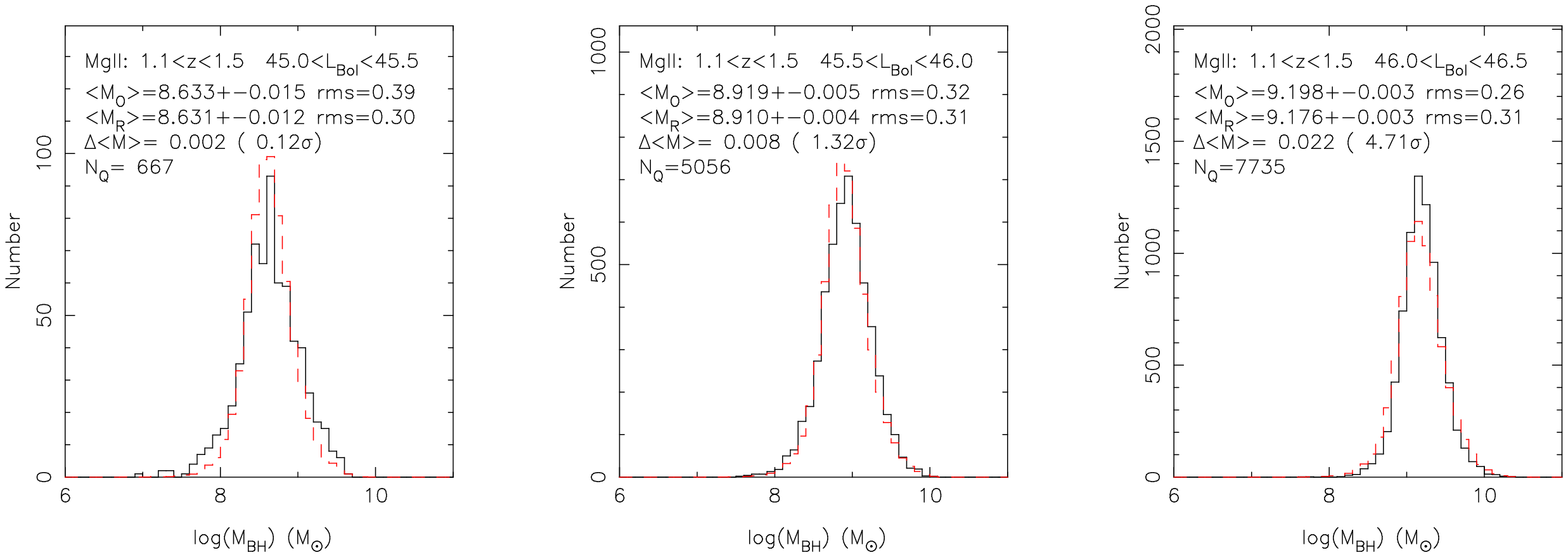}
\vspace{-6.0cm}
\caption{The distribution of black hole mass estimates in different
  redshift and bolometric luminosity intervals for \mgii.  We compare the
  original mass estimates ($M_{\rm O}$; black solid lines) to the
  randomized estimates ($M_{\rm R}$; red dashed lines).  Mean masses
  are given for both, along with the difference in the mass and its
  significance.}
\label{fig:mgiihist}
\end{figure*}

\begin{figure*}
\includegraphics[angle=270,scale=.65]{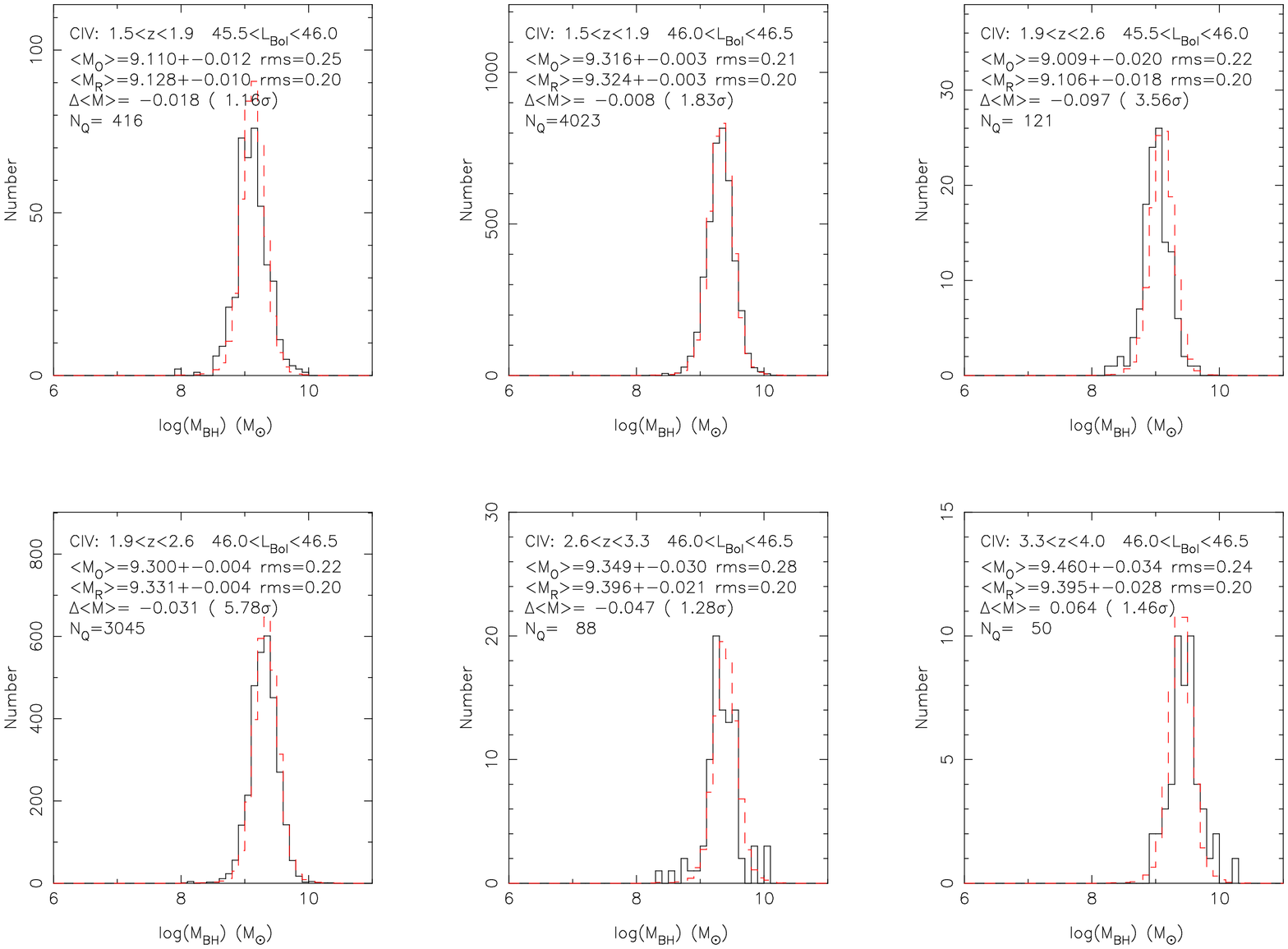}
\caption{The distribution of black hole mass estimates in different
  redshift and bolometric luminosity intervals for \civ.  We compare the
  original mass estimates ($M_{\rm O}$; black solid lines) to the
  randomized estimates ($M_{\rm R}$; red dashed lines).  Mean masses
  are given for both, along with the difference in the mass and its
  significance.}
\label{fig:civhist}
\end{figure*}

We first assess what the impact of randomization is on a single quasar
by measuring the RMS of ($M_{\rm O}$-$M_{\rm R}$).  We find RMSs of
0.54, 0.42 and 0.27 dex for \hb, \mgii\ and \civ\ respectively.  In
comparison to this the quoted uncertainties on the virial estimators
are typically 0.3-0.4 dex.  Vestergaard \& Peterson (2006) find a
scatter of 0.43 and 0.33-0.43 dex for \hb\ and \civ\ respectively,
while McLure \& Dunlop (2004) find a scatter of 0.33 dex for the
comparison between \mgii\ and \hb\ based mass estimates.  This
suggests that taking a mean velocity width to estimate a BH mass
based simply on luminosity could be a useful approach, all be it, with
some increase in the related error on BH mass.

The distribution of $\mbh$ vs. redshift is shown in Fig.
\ref{fig:hb} for the \hb, \mgii\ and
\civ\ lines.  The original distributions are shown in
black and the distribution after randomizing the emission line
velocity widths are shown in red.  In all three cases the original and
randomized black hole masses appear essentially indistinguishable.  To
examine whether there is a quantitative difference we perform a 2-D
Kolmogorov--Smirnov (KS) test (Peacock 1983).  We carry out these
tests in both the $\mbh-z$ and  $\mbh-L$ planes.  To provide a robust
estimate we generate 100 random realizations of the randomized BH
masses and then take the median $\dks$ and probability, although taking
a single realization does not significantly change our conclusions.
Table \ref{tab:ks} contains the results of the 2-D KS analysis.  For
the \hb\ line there is no significant difference between the original
and randomized  distributions of $\mbh-z$ or $\mbh-L$ (the null
hypothesis that they are drawn from the same distribution is only rejected
at the $\simeq20$\% and $\simeq10$\% level for $\mbh-z$ or $\mbh-L$
respectively).  Applying this test to the \mgii\ mass estimates we find
similar values of $\dks$, but due to the larger number of quasars the
difference between the original and randomized BH masses is now highly
significant.  Careful inspection of Fig. \ref{fig:hb} does show that
the randomized \mgii\ BH masses have a slightly broader distribution at high
redshift than the original BH masses.  This could be due to intrinsic
correlations of \mgii\ line width with redshift or luminosity (which
indeed are seen by Fine et al.\ 2008).
Alternatively, it could be driven by correlations of measurement
errors with redshift or luminosity.  Larger measurement errors will
broaden the BH mass distribution, and if signal--to--noise is 
correlated with luminosity or redshift this will cause a difference
between our original and randomized BH masses.  If we apply
corrections to the \mgii\ velocities as described in
Sec. \ref{sec:bhmass}, then the significance of the difference in the
distribution of original and randomized BH masses is reduced to only
~1-2$\sigma$.  However, to be conservative we will consider the \mgii\
mass estimates without this correction for the remainder of the paper.
For the \civ\ line the
derived $\dks$ values are again similar, and the significance of the
difference is at the 2 ($\mbh-z$) to $3\sigma$ ($\mbh-L$) level.

It might be expected that a combination of the Eddington limit and the
bright end of the BH mass function might naturally cause a narrowness
in the distribution of velocity widths.  To examine this we carry out
the same analysis as above, but this time keeping each sample
separate.  The results are listed in Table \ref{tab:ks}.  In the case
of the 2SLAQ and 2QZ samples, there is no significant difference
between the original and randomized BH masses.

Our second test is to measure the mean and RMS BH masses using each
estimator in various redshift and bolometric luminosity bins.  Again
we calculate the values for the randomized sample from 100
realizations, although we calculate the error on the mean assuming a
single random sample.  The results of this analysis are presented in
Figs. \ref{fig:hbhist}, \ref{fig:mgiihist} and \ref{fig:civhist}.  In
all cases there is a good agreement between the original ($M_{\rm O}$)
and randomized ($M_{\rm R}$) mass estimates.  For \hb\ the difference
in mean mass is always less than 0.1 dex (except for the $0.1<z<0.4$,
$46<L_{\rm Bol}<46.5$ interval which only contains 12 quasars) and the
median difference is $\simeq0.02-0.03$ dex.  Three $z-L_{\rm Bol}$
intervals have differences in the mean masses which are greater than
$2\sigma$.  The comparison of \mgii\ BH masses in
Fig. \ref{fig:mgiihist} shows similar good agreement.  The largest
difference is 0.075 dex, and the median absolute difference is only
0.01 dex.  Again, there are several intervals where the difference
between the mean masses is significant.  The \civ\ masses
(Fig. \ref{fig:civhist})  are consistent with the other emission
lines.  The maximum difference is 0.1 dex and the median is
$0.03-0.05$ dex.  Two of the intervals have significantly ($>2\sigma$)
different means.  Overall, while there are some significant
differences in the mean $\mbh$, there is impressive agreement between
the original and randomized BH masses.  Such agreement has a number of
implications, which we will discuss below.    

\section{Discussion}

The virial method makes use of the
radius--luminosity relation derived from reverberation mapping, so
that the only observables required are velocity width and luminosity.
Thus we have
\begin{equation}
M_{\rm BH}=A\times L_{\lambda}^{\alpha}\times \sigma\blr^2,
\end{equation}
where $A$ is a normalizing constant, $L_{\lambda}$ is the monochromatic
luminosity at some wavelength $\lambda$ near the emission line,
$\alpha$ is $\simeq0.5$ (e.g. Bentz et al. 2009) and $\sigma\blr$ is the width 
of the emission line.  There are multiple stages required to make the
connection between BH mass and the directly observable values.
Starting with BH mass these are:

\begin{enumerate}

\item Black hole mass ($M_{\rm BH}$) and accretion rate ($L/L_{\rm
  Edd}$) give us bolometric luminosity, $L_{\rm Bol}$.

\item $L_{\rm Bol}$ combined with bolometric corrections and
  photometric errors give us $L_{\lambda}$, {\it the first observable}.

\item $L_{\rm Bol}$ (or possibly $L_{\rm UV}$) defines the radius of
  the broad line region $R_{\rm BLR}$.

\item $M_{\rm BH}$ and $R_{\rm BLR}$, together with the assumption of
  Keplerian orbits, gives us the velocity of the broad line gas,
  $V_{\rm BLR}$.

\item $V_{\rm BLR}$ together with non-virial motion, orientation
  effects and measurement errors give us $\sigma\blr$, {\it the second
  observable}.

\end{enumerate}

An important point to note is that, apart from uncertainty in the
bolometric correction and photometric errors, scatter or errors in
each of these steps will increase the scatter in $\sigma\blr$ at a given
luminosity.  The measured uncertainties in virial relations are
typically $0.3-0.4$ dex (Vestergaard \& Peterson 2006), and these are
only statistical errors and do not include contributions from
systematic uncertainties in the reverberation mapping data, to which
the virial relations are calibrated.  The difference between the
mean BH masses measured in our original and randomized samples was at
most 0.1 dex, and typically 0.02 dex.  This is much smaller than the
uncertainties in the estimators themselves.  Indeed to highlight this
fact, we compare the SDSS \mgii\ BH masses calculated by Shen et
al. (2008) and Fine et al. (2008).  Although both use the same virial
relation calibration, the difference in method (FWHM vs. IPV) results
in a typical difference of the mean BH mass (in narrow redshift and
luminosity bins) of $\simeq0.25$ dex.  This is an order of magnitude
greater than the difference between our original and random estimates,
although it is worth noting that the line measurement methods of these
two works are somewhat different.

The above results suggest that assuming a mean velocity width and an
RMS scatter about that mean would enable us to estimate the BH mass
distribution to the same precision as we currently have using the
individual measured line widths.  This has a number of implications
for the use of virial BH masses.  One of the most common derived
quantities is the accretion efficiency $L/\ledd$.  As the velocity
width adds little information, $L/\ledd\propto L^{1-\alpha}$, where
$\alpha\simeq0.5$.  In a flux limited sample the most distant objects
will be the most luminous and so will be found to have the highest
efficiency.  

We know that quasar broad emission line properties do correlate
with various observables.  The best known case being the Baldwin
(1977) effect which correlates equivalent width with luminosity.  Fine
et al.\ (2008) finds that although the mean width of the \mgii\ line
does not depend on luminosity, the dispersion of line widths does.  There
is a significantly broader distribution of line widths at faint
luminosities.  No doubt this contributes to the small but significant
difference we do see between our original and randomized BH masses.
In contrast, Fine et al.\ (2010) find no correlation between the
dispersion in \civ\ width and luminosity (once corrected for
measurement errors).  The low overall dispersion
in line width found ($<0.1$dex at high luminosity) by both these works
naturally infers that in a flux limited sample quasar broad line
widths cannot have a large impact on BH mass estimates.  The small
scatter and other observed properties (e.g. the presence of strong
outflows indicated by broad absorption lines) may infer a problem with
our assumption that the broad line gas is virialized.  However, there
are a number of observations which suggest that virialization is at
least partially correct.  The most compelling being the correlation
between lag and line width (e.g. Peterson et al.\ 2004).  Comparisons
between broad line reverberation mapping masses and bulge velocity
dispersion shows a correlation analogous to that observed in quiescent
galaxies (Onken et al. 2004).  Added to this, there is agreement between
dynamical mass and reverberation masses for the few galaxies where
measurements have been possible (Davies et al. 2006; Onken et al. 2007;
Hicks \& Malkan 2007).

The small impact of velocity width on BH masses is due to the limited
dynamic range of velocity widths.  Under assumption that the virial
method is giving us real information concerning BH mass, the small
range in velocity widths must be in part due to joint constraints from
a steep BH mass function at high masses and the Eddington limit.  Fine
et al. (2008) examine this and find that some bias is possible, but
that selection effects cannot on their own produce the narrowness of
the velocity distribution for the most luminous quasars.  A more
robust examination would be to extend the approach of Schulze \& Wisotzki
(2010) to the larger SDSS, 2QZ and 2SLAQ samples, although this is
outside the scope of the present paper.  However, we do find that
there is no significant difference between original and randomized BH
masses when we test the fainter 2QZ and 2SLAQ samples separately.  The
dynamic range in luminosity sampled by the combined SDSS, 2QZ and
2SLAQ sample is over 6 mags for the \mgii\ line (i.e. a factor of
$\sim250$ in luminosity).  Given that $L/\ledd\sim L^{0.5}$ this
corresponds to a factor of $\sim16$ change in $L/\ledd$.  It is rather
remarkable that over this range there is no change the mean velocity
width in the \mgii\ line.

Clearly, if a sub-sample is selected on the basis
of line width, then randomizing those velocities
with the rest of the population will have significant impact on the
derived BH masses.  For example, narrow--line Seyfert 1s (NLS1s) are
rare in any flux limited sample and have been
considered to have abnormally low BH masses.  Such randomization as we
carry out will clearly change dramatically the mass estimates of these
objects.  However, recent work has suggested that NLS1s do not have
low masses but that their narrow lines are caused by orientation
(Decarli et al.\ 2008) and/or radiation pressure (Marconi et al.\
2008).  Such effects should increase the scatter in velocity width
for a given black hole mass.  However, the impact of this is limited
by the small observed scatter as a function of luminosity.

Specific quasar sub-samples have been shown to have different virial BH
mass distributions.  One particular example is radio--loud or
radio--detected quasars.  Jarvis \& McLure (2006) show that a sample
of radio quiet quasars has narrower broad lines than a matched
radio--loud sample.  This demonstrates that the difference between
radio--loud and radio--quiet quasars is not due to selection biases.
However, such biases do exist.  The fraction of radio--detected
quasars increases with luminosity (e.g. Jiang et al.\ 2007), as one
would expect in a flux limited sample, so any radio-loud sample will
have higher estimates for BH mass even before accounting for the
intrinsic property differences. 


A positive outcome of the weak impact of quasar broad-line width on
black hole mass estimates is that a reasonable mass estimate can be
derived without even measuring the width of quasar emission lines!
With improved multi-band photometry (e.g. including near- and mid-IR
data, Richards et al.\ 2009) high quality, faint, photometric samples
of quasars can be obtained, with reasonable photometric redshifts.  It
would then be possible to derive a quasar BH mass function without
obtaining spectra.  However, inferences from such analyses would be
limited.  Basing BH masses simply on observed luminosity does not 
allow us to investigate accretion rate variations.  

A clear route to improve our current position it to attempt better
calibration of the virial mass relations, and in particular determine
how other parameters influence line width (e.g. orientation, radiation
pressure).  A key question, which has yet to be answered, is whether
there is any reason for the coincidence that accretion efficiency
scales with luminosity (as $L/\ledd\sim L^{0.5}$) such that quasar broad
line widths are constant with luminosity.    






\acknowledgments

SMC acknowledges the support of an Australian Research Council QEII
Fellowship and an J G Russell Award from the Australian Academy of
Science.  We acknowledge the organizers and attendees of IAU 267 in
Rio de Janeiro for discussions which triggered this work.  We
acknowledge Stephen Fine and Yue Shen for making their quasar line fits
public and also acknowledge Stephen Fine and Gordon Richards for
useful conversations.  We thank the anonymous referee for useful
comments.

{\it Facilities:} \facility{AAT}, \facility{SDSS}.




\clearpage

\end{document}